# Anomaly in the HTSC system $(BiPb)_2Sr_2Ca_2Cu_3O_{10-\delta}$ at the temperature 203.7K

I. Akhvlediani[*,1], T. Kalabegishvili[**,1,4], N. Kekelidze[2], N. Margiani[3] and N. Papunashvili[3]

[1] Iv. Javakhishvili Tbilisi State University, E. Andronikashvili Institute of Physics, 6 Tamarashvili str., 0186 Tbilisi, Georgia
[2] Iv.Javakhishvili Tbilisi State University, 0179 Tbilisi, Georgia
[3] Georgian Technical University, Institute of Cybernetics, 0175 Tbilisi, Georgia
[4] Ilia State University, Institute of Applied Physics, 0162 Tbilisi, Georgia

[*] Corresponding author: e-mail ioseb.akhvlediani@tsu.ge, Phone: (995) 558 542 836, Fax: (995 32) 39 14 94
[**] e-mail: tamaz.kalabegishvili@tsu.ge, Phone: (995) 593 225 100, Fax (995 32) 39 14 94

## Abstract

Broadening of the electron paramagnetic resonance (EPR) line of the paramagnetic probe decorated on the superconducting system surface was detected at the temperature $T>T_c$. The source of the anomaly and the conditions of its emergence were revealed. The analysis of the experimental results with taking into account the literature data serves as the basis to consider that the observed broadening of the EPR line of the paramagnetic probe at 203.7K resulted from the transition of the system under study into the superconducting state.

Investigation of anomalous phenomena in high-temperature superconductors (HTSC) in the normal state ($T>T_c$) is quite actual. Revealing the anomalies, determination of their source and understanding of the nature of anomalies of HTSC systems is important for elucidation of the mechanism their superconductivity.

In this work we used HTSC systems $Bi_{1.7}Pb_{0.3}Sr_2Ca_2Cu_3O_{10-\delta}$ (2223) and $YBa_2Cu_3O_{7-\delta}$ (123) as the object of study. The samples were prepared by a standard method of hard-phase synthesis of chemically pure powders [1]. For investigation as of the superconductivity transition so of anomalies in the normal state, the method of electron paramagnetic resonance (EPR) was used. To this purpose the surface of the sample was decorated with a paramagnetic probe. In our case we used diphenyl-picryl-hydrazyl (DPPH) radical and observed the change in the EPR linewidth for this radical with temperature both below and above the critical temperature of the superconductivity transition. For decorating DPPH on the sample surface, we placed the sample into the $10^{-2}$ M solution of DPPH in acetone and then dried it in the air. For studying the temperature dependence of the EPR linewidth, we placed a double quartz ampoule containing the sample, a heating element and a thermocouple into a quartz dewar and filled the dewar up with liquid nitrogen. The EPR measurements were carried out with the help of spectrometer RE 1306 operating at the frequency 9.4 GHz (X-band) and modulation 100 kHz. The amplitude of modulation and the power level were selected in order to exclude the broadening and saturation of the EPR line. The equipment allowed us to measure the EPR spectra over the temperature range from 77 to 300K. The heating element was wound on the inner ampoule in order to exclude a decrease in the quality factor of the resonator [2]. Having measured the spectrum at room temperature and then cooling the sample rapidly, we measured the spectrum at the liquid nitrogen temperature. The following values of the temperature were obtained by varying the value of the current passing through the heating element. The temperature stability was equal to ±0.5K. In the Figure (a) is shown the temperature dependence of the EPR linewidth of the paramagnetic probe, decorated on the surface of the high-$T_c$ superconductor (2223).

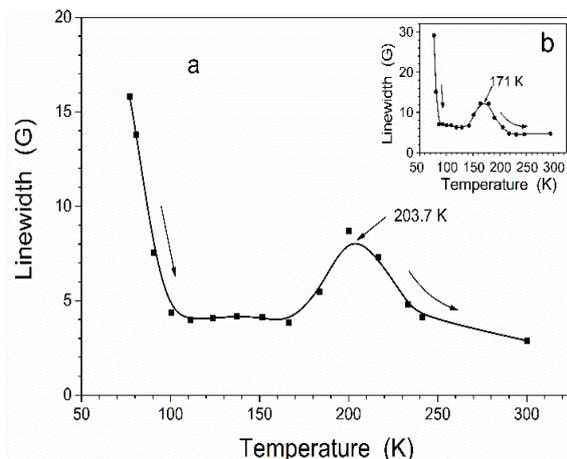

Figure. Temperature dependence of the EPR linewidth for DPPH decorated on the surface of HTSC samples: a – for $Bi_{1.7}Pb_{0.3}Sr_2Ca_2Cu_3O_{10-\delta}$; b – for $YBa_2Cu_3O_{7-\delta}$

As it is seen from the Figure, at the temperature of 203.7K, there is a clear peak of the EPR linewidth. Earlier we observed similar peaks of the EPR linewidth at the temperatures 155K; 175K; 185K; 190K; 197K (the cause of so different temperatures of anomalies has not been established yet). As it is seen from the Figure, to the left of the peak, the EPR linewidth does not depend on the temperature in the interval from 150 to 107K. A considerable increase in the linewidth is observed below $T_c$, which is associated with the transition of the sample into the superconducting state. Herewith the Abrikosov vortices arise and they change sharply the homogeneity of the magnetic field, which accounts for the increase in the EPR linewidth. The checking experiments carried out for polycrystalline DPPH and also for DPPH decorated on non-superconducting metallic copper exactly of the same shape and dimensions as the studied samples (2223) showed that there was no change in the EPR linewidth of the paramagnetic probe over the entire temperature range [3,4]. Thus, it is obvious that the peak at 203.7 is associated only with HTSC sample $Bi_{1.7}Pb_{0.3}Sr_2Ca_2Cu_3O_{10-\delta}$, which is in the normal state.

To understand what the linewidth anomaly source was at 203.7 K, we carried out the same temperature measurements for the $YBa_2Cu_3O_{7-\delta}$ system. In this case as well the experiments showed broadening of the line in the normal state but at 171K (Figure (b)). The identity of the behavior of both HTSC systems in the normal state (appearance of anomalous peaks) points at that it is connected with that common link which is present in both systems, namely with the $CuO_2$ plane. The availability of the peak at 203.7K may be associated with the superconducting transition or the magnetic phase transition, i.e. with the formation of a new magnetic phase. In order to find out whether this peak is connected with superconductivity, we carried out the experiments with the samples that did not possess the superconducting properties. These experiments showed the total absence of any peak over the entire temperature range from 77 to 300K. Thus, based on the experimental results, we can say that, if there is no superconductivity there is no anomaly in the temperature range $T>T_c$, and, if there is superconductivity, there emerges an anomaly at $T>T_c$. It is important to note that the given anomaly appears only when the cooled sample is heated. In this case in bismuth systems an increase in the lattice parameter takes place (see [5], Fig.4). In paper [6] it was shown that the deformation increase of the lattice parameter along the axis *c* increases the value of critical temperature. The analysis of the mentioned experimental results with taking into account the facts described in [5, 6] serves as the basis to believe that the anomaly (broadening of the EPR line) at 203.7K may result from the HTSC system transition into the superconducting state.

Other anomalies of copper-containing systems will be reported later.

## References

1. N. Margiani, G. Mumladze, Z. Adamia, N. Papunashvili, D. Dzanashvili. J. Supercond. Nov. Magn. 28, pp.499-502, 2015.
2. T. Kalabegishvili, J. Aneli, I. Akhvlediani, J. Chigvinadze. Author Certificate GE P 2005 3440 B, GE, Tbilisi, 2005 (In Georgian).
3. J. Chigvinadze, J.Acrivos, I.Akhvlediani, M.Chubabria, T.Kalabigishvili, T.Sanadze. Physics Letters A **373**, 8-9, pp.874-878, 2009.
4. J. Chigvinadze, I, Akhvlediani, T. Kalabegishvili, T. Sanadze, Nano Studies, pp.245-249, 2010.
5. S. Pryanichnikov, S. Titova, G. Kaljuzhnaia, J. Gorina, JETP, **134**, pp.89-94, 2008.
6. J.P. Locquet et.al, Nature, **394**, p.453, 1998.